\documentclass[runningheads]{llncs}
\usepackage{amsmath,amsfonts}
\usepackage{array}
\usepackage{url}
\usepackage{verbatim}
\usepackage{graphicx}
\usepackage{cite}
\usepackage[T1]{fontenc}
\usepackage{multirow} 
\usepackage{hyperref} 
\usepackage{xcolor} 
\usepackage{subcaption} 
\usepackage{pifont}
\usepackage{tcolorbox}
\usepackage{xspace}
\usepackage{tabularx}
\usepackage{orcidlink}
\usepackage{amssymb}
\usepackage{colortbl} 
\usepackage{adjustbox}
\usepackage{enumitem}

\newcolumntype{Y}{>{\centering\arraybackslash}X}

\hypersetup{
    colorlinks=true,
    linkcolor=blue,
    filecolor=magenta,      
    urlcolor=blue,
    pdftitle={Pricings Report},
    pdfpagemode=FullScreen,
}

\newcommand{\chunk}[2]{}

\newcommand{\code}[1]{\texttt{#1}}

\newcommand{\modelNameNoSpace}{HORIZON}
\newcommand{\modelName}{\modelNameNoSpace\xspace}

\newcommand{\cmark}{\textcolor{black}{\ding{51}}}
\newcommand{\xmark}{\textcolor{black}{\ding{55}}}
\newcommand{\smark}{\textcolor{black}{$\sim$}} 

\newcounter{rqAnswerCounter}

\begin{document}

\title{HORIZON: a Classification and Comparison Framework for Pricing-driven Feature Toggling}
\titlerunning{HORIZON: a Comparison Framework for Pricing-driven Feature Toggling}

\author{Alejandro García-Fernández\inst{1}\orcidlink{0009-0000-0353-8891} \and
José Antonio Parejo\inst{1}\orcidlink{0000-0002-4708-4606} \and
Antonio Ruiz-Cortés\inst{1}\orcidlink{0000-0001-9827-1834}}
\authorrunning{A. García-Fernández et al.}

\institute{SCORE Lab, I3US Institute, Universidad de Sevilla, Espa\~na \\
\email{\{agarcia29,japarejo,aruiz\}@us.es}}
\maketitle              








\begin{abstract}
Software as a Service (SaaS) has seen rapid growth in recent years, thanks to its ability to adapt to diverse user needs through subscription-based models. 
%
However, as pricing models enhance the customization of subscriptions, managing the associated constraints within a system's codebase becomes increasingly challenging.
In response, Pricing-driven Development and Operation has emerged to integrate pricing considerations across the software lifecycle. Among its most challenging objectives is regulating feature access according to users’ subscriptions ---a process that requires managing a multitude of conditions throughout the system’s codebase.
%
Feature toggles have traditionally been employed to manage dynamic system behavior, but their application to pricing-driven constraints presents unique challenges. When used to enforce subscription-based restrictions, toggles must adapt ---among other factors--- to individual user's use of features, ensuring that subscription limits are not exceeded.
%
Despite the increasing significance of this problem, current industrial solutions lack explicit support for pricing-driven feature toggling, and existing academic contributions remain constrained to specific architectures.
This paper contributes to fill this gap by introducing \modelName, a classification and comparison framework for feature toggling tools tailored to pricing-driven environments. Its utility is demonstrated by categorizing the solutions identified in the literature as promising for such environments, revealing both their strengths and limitations, and thereby pinpointing critical avenues for improvement. In doing so, \modelName not only provides a comprehensive view of the current landscape but also lays the groundwork for a focused research agenda, guiding the development of more robust and adaptable solutions for streamlining SaaS development and operations driven by pricings.

\end{abstract}

\section{Introduction}
\label{sec:introduction}

Software as a Service (SaaS) has rapidly gained popularity in recent years \cite{Jiang2009}, driven by its flexibility to adapt to diverse user needs through subscription-based models. However, this adaptability has led to an exponential increase in configuration complexity, with modern pricing models evolving from offering a few options to thousands of potential configurations \cite{ICSOC2024}. For example, Salesforce’s November 2019 pricing offered up to 10 different configurations, while its July 2024 version allowed for over 12,000 unique configurations (as can be seen \href{http://sphere.score.us.es/pricings/card?name=Salesforce}{here}). This expansion presents significant challenges for the development and operation of SaaS-based Information Systems (IS), as manual management of these configurations becomes a demanding, error-prone, and time-consuming task.

The paradigm of \emph{Pricing-driven Development and Operation}\footnote{For brevity, we may also refer to this concept as Pricing-driven SaaS DevOps.} has emerged to address these issues, which in the end share the same common need: optimizing processes impacted by pricing ---and, consequently, by market and competitors--- throughout the SaaS lifecycle, such as pricing design or feature access control based on user subscriptions. In particular, managing access in accordance with pricing constraints is particularly challenging, as it requires efficiently handling numerous conditional checks dispersed throughout the codebase to ensure that each user accesses only the features corresponding to their subscription.


Feature toggles ---a technology that enables dynamic changes in system behavior without modifying underlying code \cite{FOWLER2023}--- offer a promising solution to this challenge. However, given the variability of pricings \cite{ICSOC2024}, managing them manually quickly becomes unwieldy and error-prone. Although industrial solutions exist to centralize toggle configurations and simplify their management, they were originally designed for A/B testing and canary releases rather than for dynamically enabling or disabling features based on subscriptions, i.e. pricing-driven feature toggling. This limitation largely stems from the historical absence of standardized pricing metamodels \cite{CAISEFORUM2024}, forcing providers with complex configurations to develop custom, ad-hoc solutions. Notably, the only approach supporting pricing-driven feature toggling ---Pricing4SaaS \cite{ICWE_DEMO2024}--- has been validated exclusively in academic settings and for client-server architectures. 

Hence, to support progress towards this vision of pricing-driven feature toggling, i.e. to provide systems with the ability to adapt to pricing changes autonomously, this paper introduces \modelName, a classification and comparison framework for feature toggling tools tailored to pricing-driven environments that aims to guide future research and industrial efforts towards new solutions in this area. To demonstrate its applicability, the framework was utilized to extend the experimental analysis of feature toggling tools reported in \cite{ICSOC2024}, with the aim of pinpointing the functionalities these solutions currently lack for adequately supporting pricing-driven feature toggling. Consequently, \modelName not only serves as a benchmark for evaluating and comparing existing approaches but also acts as a practical resource for developers seeking to enhance the management and automation of their feature toggling solutions ---whether in ad-hoc setups or integrated with third-party tools.

The remainder of this paper is structured as follows: Section \ref{sec:background} presents the core concepts of pricings and feature toggles, and frames our contribution in the literature. Next, Section \ref{sec:proposal} introduces \modelName and Section \ref{sec:validation} demonstrates its application in assessing the capabilities of feature toggling tools within the context of pricing-driven DevOps. Finally, 
Section \ref{sec:conclusions} showcases the conclusions.

\section{Related Work}
\label{sec:background}

\subsection{SaaS Pricings}
\label{sec:background:pricing}

A pricing is a part of a SaaS customer agreement \cite{garcia2021flexible}. It structures the \textit{features} of a service ---defined as the distinctive characteristics whose presence/absence may guide a user’s decision towards a particular subscription \cite{JCIS2024LONG}--- into \textit{plans} and \textit{add-ons} to control users' access to such features. While users can only subscribe to one of the available plans, they can subscribe to as many add-ons as they want, as long as they are available for the contracted plan. As can be noted, in this domain, a feature encompasses both \textit{functional features}, which constitute the core software product, and \textit{extra-functional features}, which, while external to the product itself, enhance its perceived value (e.g., support, SLA guarantees, etc). By including both types of features, pricings capture a more comprehensive view of what influences customer value, aligning the service offering with the overall customer agreement.

\begin{figure}[htb]
    \centering
    \includegraphics[width=\textwidth]{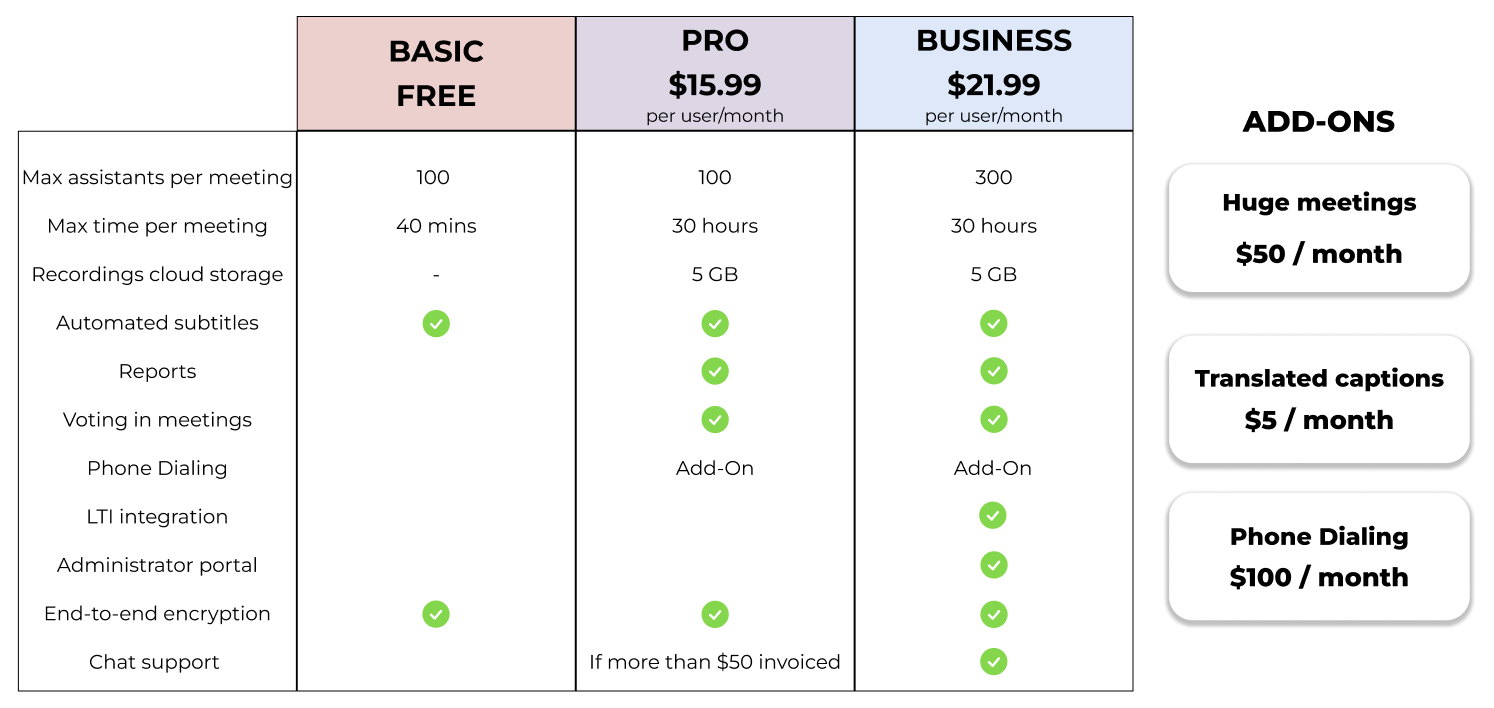}
    \caption{Excerpt of Zoom's pricing with 11 features, three plans and three add-ons.}
    \label{fig:zoomPricing}
\end{figure}

To illustrate these concepts, consider the real-world example of \href{https://zoom.us/pricing}{Zoom}. This cloud-based video conferencing service enables users to conduct virtual meetings and, optionally, record sessions for later viewing. An excerpt of its pricing is presented in Fig. \ref{fig:zoomPricing}.\footnote{Pricing entries that impose or extend limits on meetings are considered usage limits rather than individual features—the overarching feature in this case is “meetings”.} In this example, nine features are managed through plans, one feature is associated with an add-on (``translated captions''), and one is governed by both (``phone dialing''). The pricing also enforces usage limits on the ``meetings'' feature (e.g., maximum assistants per meeting and maximum meeting duration) meaning that although the feature is available in all plans, the extent of their usage differs ---higher-priced plans offer higher limits.

The paradigm of \emph{Pricing-driven Development and Operation} has emerged to address the challenges associated with pricing management tasks. However, despite growing interest, there has been limited progress in this area. The Pricing4SaaS model, introduced as a formal metamodel, represents one of the first generalized attempts to standardize pricings \cite{CAISEFORUM2024}. It defines pricings as a combination of \textit{plans} and \textit{add-ons} that regulate access to a set of features whose usage may be limited by usage limits. Additionally, its YAML-based serialization, Pricing2Yaml, has served as the initial step towards automated tooling for Pricing-driven DevOps \cite{ICWE_DEMO2024} and thus iPricings, i.e. machine-oriented pricings \cite{AI4PRICING}.\footnote{You can find the Pricing2Yaml serialization of Zoom's pricing \href{http://sphere.score.us.es/editor?pricing=M4QxwORBbBTAuABALQPaugKGATwHYAuIAHgGqwBOwAlqnkgOQBMAdAIwOYBulNdSTAAxMALAFo2bMSMwBjCrBAFYAEwCCBRkNESpgkZ1kBXCgryycSAKoBlACKYAZooInYweJkSI4sAtTwAcw8vb0QuEAAbI1gAFRwABwREACEAeTSAGQBRNQhQ7xVYRxAjSIJSKJikAgoYgsQCROS7NIBZNQBJfO9ZSNQjFQAlWFlUChUA4M8w8Kq45qR0rNyesKKSsor5pBLI4FgGpqSkVo7u0NKCDCVVGyMAI38CSPcZsIjohZPUjJy8hobUrlSpfGp1Q6zY7JNRWWLtNSxTppNbeK43fx0eI-dKxUIKBLjAghWafGLY5LLf6oxBArag6qIPYHI6LRDdABiaSGHSRKNCXFQ-iCnTwbVgfimJI+8wpSz+q0BxWB2zBTKiLKhbLOXTWCQAFnRYHZqFEpu9vGTvpSFQDZnSQTt1ftIWFoUhurFsgBxIaI5E0gLKQIUJS0PByxDZAAaXqGEDUmQA+nZsqROgBhbKhcrUUXB0OY+gNK2RqmK+3K+lO5mu7zu9kQL2+-382ZB2AhsNYtkdIYAaWysQACpk1FnLipoAFqMBakpxsOiVELXMvmXbTSHarGbXWT8Ognvdk2tkm6FYHgVGJrmJLypsuYKIki6vS2zy3b1lXHWravUtR+b0rDUP0m2ybNen1JR7gSQkKE0EtZTZL1YyVTZf0ZAAiLD92SGwrGHYduTxetQzwYBIluFQMxABIi2lS1kJxTd0JVBlkj3QCYThBE+RpdFoG7CMPzSPEjFAQJYEyahp2Jd4hOINRgBoOcQEIYBh0ocVJSCN9mOSCArFPIZMzY6s1SYBojDwahNEQCTYAAemgOgCH1PCkCGM9sgAdTUFIcgaMwigoJAzwCnIHFmSIAgAa1UDkXDcRiwjEHwJWFYJQkU2JZNgLSKB0rL9PXNkjJMszKwwndkhEQRrNs+zpzwFzMqmTzEAgFEk28iA-MiyCwhCyhwoTQLsmisJYrwBKVCSpQUtXbx0t8LLgHxUZxkmIJgAzfpBhsa5Qyk0ryXK4zslMjNzMw5IGtmGy7KQb0Uk67qIF6nz-ImhoZrmhbXAUVKVsQPoBmGLaJilTAEioij3hSNQbEzVcimAeRqHo8MlnAahZEQOH1IaBIKHx+7GuehyDgoFy3I82ZnEW4HlrCWRoIIWD4MQ2ZeZlNUuLCCSQCkmS5I8RA8DKSJQmHIY0jR9xMex-hEDUFQInMVRCfhkmydkZI2AAVhYABOU3KfsxzadcwgGbCJmgbeBpegOyGxmh3bWaYv8IRdxACSJEH+cZf86w+IUplFYqpW9tdQ79vn2Zgow4KJOOrSQLDOkcHxxlgRpoLwRAABIjcERAAkFcmVFwx7JOk2S7OD3L8sKmO9P9n3GTYAAOQQHt5hQPZ24J9oho7xhFhAu-j5IjdCFJbG6bIbBsRWMbJlX6DVjX1INlQdeJ2ZSfJgQ2DNi3HqapBrbpu2Gkdpb-fBwYRhH2PZ8zxpE6H2BueDt3ZIYd-aCiytHdqXsv5OhAXzXM+ZOyFhxtA32AFeYgCnDOOchZFzLkiBnGBv9ZjJ05qnbmBDUHh2FqLJu8l-aKWUqpIgGl26QMCBQxkABmAe9CSB5TgKw3S7CUE937oPWYw9tpSnHodY608OHz0wBglQaQEahH1EYKSHdpgNBABEagVEHivA5OMVm6UkYoxunzdKcs0j+3Mcvfqa89Zn0QOXS2SB77uWsg3MWzdWYMJUrOZhxJBElREYbAeg95wUSosoGidEGKrj0SAAxIAjGwBMWFexqRkZVV5jY+WOSl4oycTYFxBskAL2vlTLx9tvBPxZv7GJlFqK0W3oAue4I0EGiNCaM0ndZgpLSRkrJZjEC2OKY41e5ST760ieIhyN9EB1MfslJpfNel4GNKaGawi+ZAO6YcIAA}{here}}


\subsection{Pricing-driven Feature Toggles}
\label{sec:background:featureToggling}

A recent large-scale analysis \cite{ICSOC2024} modeled and compared up to 162 pricings (from 30 SaaS providers, spanning multiple years), revealing how the \textit{configuration space} of pricings, i.e. the number of potential different subscriptions that can be created from a pricing,\footnote{For instance, Zoom's configuration space is 20} tends to grow exponentially over time ---driven primarily by the addition of add-ons. This growth not only diversifies the range of available subscriptions and increases customization for users, but also complicates the development and operational management of features within the software.

\textit{Feature toggles} are a promising approach for managing such complexity.
They can be defined as a software development technique that allows features to be dynamically enabled or disabled without modifying the code \cite{FOWLER2023}. In a nutshell, this behavior is implemented by using boolean expressions whose values can be assigned, or modified, at runtime, thus providing dynamic evaluations. Fig. \ref{fig:Feature-toggles-criteria} illustrates the simplest version of a feature toggle in the source code of a hypothetical implementation of Zoom. This toggle evaluates the feature ``Records'', enabling or disabling its web component based on the user's plan. As shown, the evaluation of the conditional block is hard-coded, but some values depend on dynamic data, e.g. ``\textit{userPlan}'', which is retrieved using the ``\textit{fetchUserPlan}'' function from the toggle context, an external source (e.g. a database) that contains the data needed for the evaluation.

Building on this concept, a widely recognized classification of feature toggles distinguishes a type known for its high dynamism and extended lifespan ---namely, permission toggles \cite{FOWLER2023}. In pricing-driven SaaS environments, permission toggles that are used to provide each user with a different configuration of the service at runtime according to their subscription have been referred to as pricing-driven feature toggles \cite{ICSOC2024}.
By mapping iPricings into these toggles, developers can manage multiple configurations within a single deployment instance of the application and dynamically assign the appropriate features to each user based on their subscription tier, fully leveraging the multi-tenant potential of SaaS \cite{GHADDAR2012}. Unfortunately, this flexibility comes at a cost. The integration of feature toggles, especially pricing-driven ones, introduces operational complexity, as each toggle adds new dimensions of variability to the service. Moreover, their management tends to generate ``one of the worst kinds of technical debt'' over time \cite{TERNAVA2022}, while also transforming testing into a combinatorial challenge \cite{Rahman2016}.

\subsection{Current Support for Pricing-driven Feature Toggling}
\label{sec:background:pricingDrivenFeatureToggling}

Modern feature toggle management solutions address the inherent complexity of toggles by shifting their logic from application code to external systems.  This abstraction eliminates the need for hard-coded evaluations, allowing feature availability to be determined dynamically based on configurable rules. In practice, runtime evaluations in an application might be reduced to a simple method call such as \textit{isFeatureAvailable(featureId)}, where the actual logic governing the feature’s activation is managed externally.

At their core, feature toggling solutions share a similar procedure to manage the evaluation of features. First, they introduce a different notion of \textit{feature}. In their context, a feature is a toggleable unit of functionality, to which \textit{rules} (the logic dictating whether the feature is enabled or not) are attached in order to perform an \textit{evaluation} with a given \textit{context} (the data used to evaluate the rule, such as a user’s plan or subscription in a pricing-driven environment) for a specific user or \textit{segment} (group of users that share a common context).


\begin{figure}[htb]
    \centering
    \includegraphics[width=\linewidth]{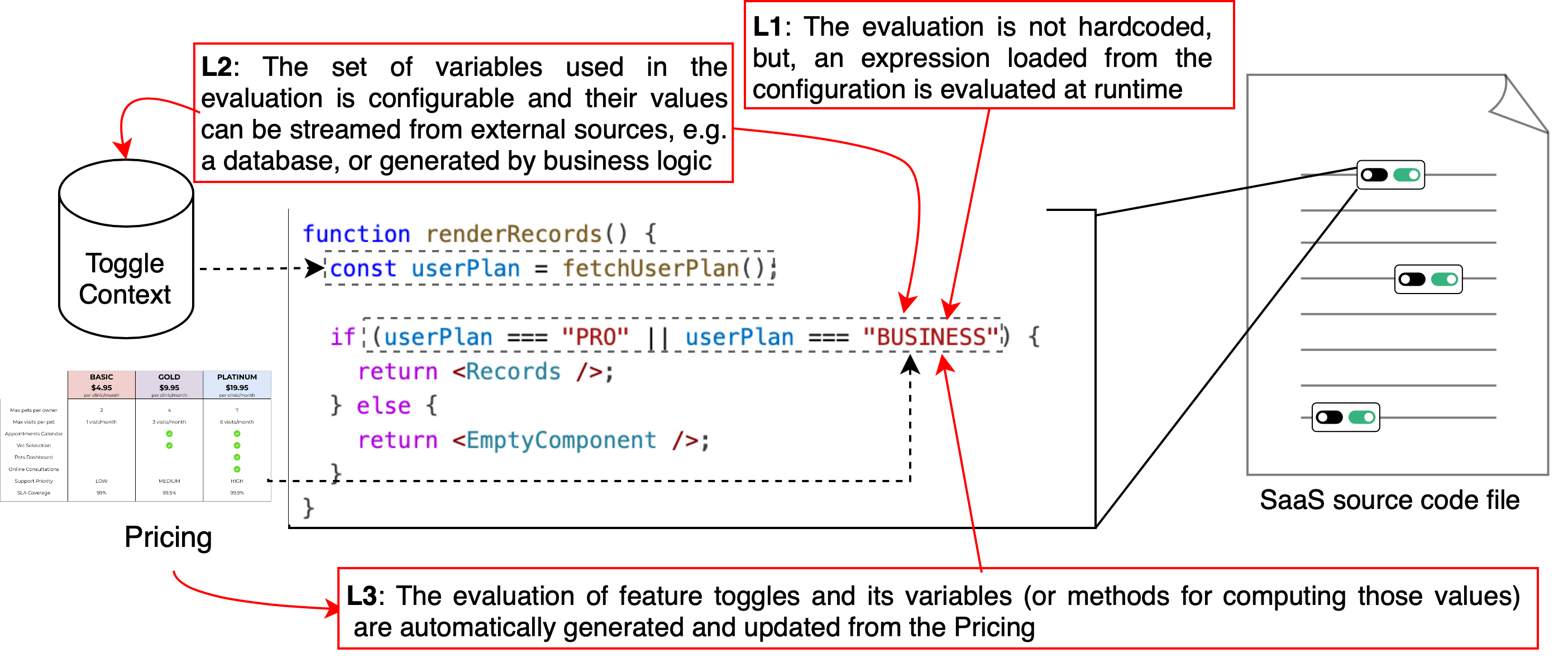}
    \caption{Feature toggle evaluating a user’s plan before enabling a feature}
    \label{fig:Feature-toggles-criteria}
\end{figure}

Unfortunately, as demonstrated by \cite{ICSOC2024}, there is a widening gap between the configurability offered by pricings and the actual capabilities of feature toggling solutions, which struggle to manage the increasing variability. To better understand these limitations, the authors propose a classification system to measure the capacity of feature toggling solutions to efficiently support pricing-driven feature toggles, introducing three distinct levels (see Fig. \ref{fig:Feature-toggles-criteria}):\\
\emph{L1: Configuration-based toggling:} feature toggling logic is externalized into a configuration file or engine that can be modified at runtime, but developers must manually update it whenever the pricing changes.\\    
\emph{L2: Dynamic and extensible contextual evaluation:} in addition to L1, rules can dynamically retrieve context variables (e.g., a user’s subscription) from external sources at runtime, eliminating the need for predefined values.\\ 
\emph{L3: Pricing-aware evaluation:} features and rules are automatically generated from an iPricing, such as a serialization in Pricing2Yaml, allowing the system to seamlessly adapt to changes in its pricing without manual intervention.


While these levels provide a high-level assessment of the essential requirements for pricing-driven feature toggling ---where each level serves as a prerequisite for the next--- they fall short in identifying specific areas for improvement. In contrast, \modelName adopts a more constructive, capability-based approach. It outlines a set of characteristics that are necessary for fully supporting pricing-driven feature toggling automation in practice. Such guidelines can then be used either to assess the capabilities of a tool or to provide a clear roadmap for enhancing its support for pricing-driven feature toggling.


\section{\modelName: Guiding Pricing-Driven Solutions}
\label{sec:proposal}

As discussed in Section \ref{sec:background:pricingDrivenFeatureToggling}, there is a gap between the evolution of SaaS pricings and the ability of industrial feature toggling solutions ---such as \href{https://github.com/Unleash/unleash}{Unleash} and \href{https://www.devcycle.com}{DevCycle}--- to handle pricing-derived variability effectively \cite{ICSOC2024}. These tools have yet to achieve level 3 support for pricing-driven feature toggling, meaning they lack the automation needed to keep pace with the increasing complexity of pricing models. As a result, maintaining pricing-driven toggles in these solutions relies heavily on manual intervention, which tend to become unsustainable as pricings continue to grow and evolve.

As a step toward closing this gap, we introduce \modelName, a classification and comparison framework for feature toggling tools tailored to pricing-driven environments. Embracing a constructive, capability-based approach, \modelName outlines the essential features required to fully support the vision of pricing-driven feature toggling, i.e. to provide systems with the ability to adapt to pricing changes autonomously.
To facilitate its practical application, the framework divides such functionalities into five distinct areas, 
each one focusing on a specific aspect of feature toggling systems that contributes to the overall objective:

\begin{itemize}
    \item \textbf{Feature management:} establishes the foundation for defining and maintaining feature toggles, ensuring the system has the necessary mechanisms to structure toggles properly before their evaluation.

    \item \textbf{Evaluation configuration:} focuses on the logic governing feature activation, i.e. it defines the expected capabilities of the expressions that determine whether a given feature is active for a particular user or context.

    \item \textbf{Feature evaluation:} addresses the mechanisms for executing these evaluations, such as retrieving the status of every feature for a specified user, or computing a single feature’s availability within a specific context.

    \item \textbf{Integration:} describes how the toggling system connects to the external applications that rely on its engine to manage the availability of their features.

    \item \textbf{Pricing-driven automation:} details model-driven approaches for automatically generating feature toggle structures based on a pricing model.
\end{itemize}

Within this structure, \modelName specifies a series of capabilities and classifies them as either required or optional to achieve full support and automation for pricing-driven feature toggling. Any solution that implements all the required capabilities is deemed to fulfill the framework’s objective, even if it does not encompass all the features of an ideal pricing-driven feature toggling engine. 

In the following, we showcase each of \modelNameNoSpace's areas in detail, outlining their key functionalities and explaining how they contribute to building a solution capable of managing pricing-derived variability in SaaS applications.

\subsubsection{Feature Management} is the foundation of any feature toggling system, enabling applications to decouple feature definitions from their codebase and store them externally—whether in a separate configuration file, a database, or a remote service. This separation is a fundamental aspect for leveraging feature toggling to its full potential, as it allows teams to control feature availability dynamically without modifying application logic or redeploying the system.

However, in a pricing-driven feature toggling context, this level of decoupling is not sufficient. Unlike traditional feature toggles, which often rely on static boolean values (e.g., true for enabled, false for disabled), pricing-driven feature toggles require rule-based evaluations, where feature availability is determined dynamically through configurable evaluation expressions. These expressions define the conditions under which a feature is enabled or restricted, allowing toggles to respond to complex pricing constraints rather than simple binary states. Moreover, to achieve level 2 support for pricing-driven feature toggling (see Section \ref{sec:background:pricingDrivenFeatureToggling}), these evaluation expressions must be capable of integrating external data sources, meaning that feature availability should not depend solely on pre-configured values but instead be dynamically assessed based on real-time information such as subscription tiers, quota usage, or external database queries.

\begin{figure}
    \centering
    \includegraphics[width=\linewidth]{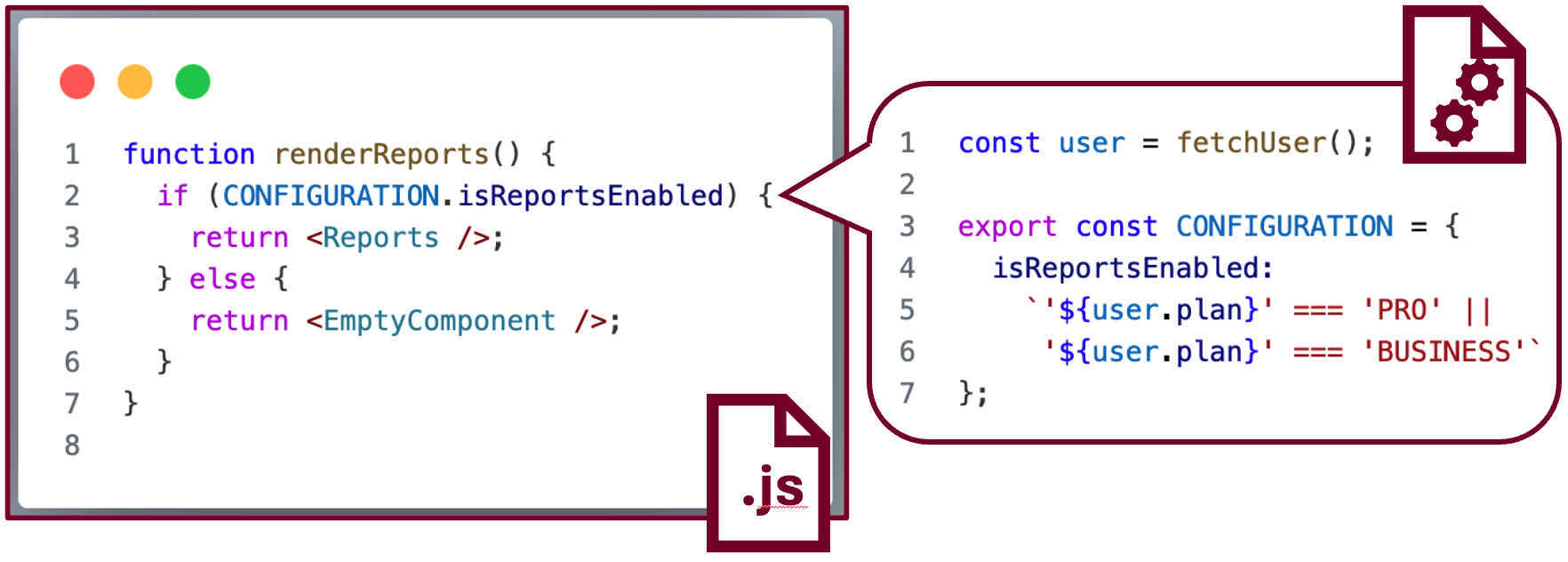}
    \caption{Definition of feature toggle evaluation in an external source}
    \label{fig:evaluationsInFile}
\end{figure}

To meet these requirements, \modelName defines the following characteristics as required for an ideal pricing-driven feature toggling management system:

\begin{itemize}[align=left, leftmargin=0pt, labelindent=\parindent,
listparindent=\parindent, labelwidth=0pt, itemindent=!]
    \item \emph{Feature CRUD operations}: The system must provide full lifecycle management of feature toggles, allowing them to be dynamically created, retrieved, updated, and deleted. For example, in the case of Zoom’s ``Reports'' feature ---available only for PRO and BUSINESS plans--- the system should be able to define (CREATE) the toggle when introducing the feature, retrieve (READ) its properties to verify in which environments it is enabled and what rules govern its activation, modify (UPDATE) its rule set if access is later restricted to BUSINESS users, and remove (DELETE) it when the feature is deprecated. Importantly, modifying a feature in this context refers to structural changes, such as adjusting the environments where it applies, updating its identifier, or reassigning the rules involved in its evaluation. This does not include altering the internal logic of its evaluation expressions, which is considered separately, under the Evaluation Configuration area.

    \item \emph{Rule CRUD operations}: The system must support full lifecycle management of the rules that govern feature evaluation, enabling their creation, retrieval, modification, and deletion as needed. For example, to manage Zoom’s ``Max Time per Meeting'' usage limit (see Fig. \ref{fig:zoomPricing}), a new rule enforcing this limit can be CREATED, retrieved (READ) for debugging or administrative review, reassigned (UPDATED) to the ``recordings'' feature if Zoom decides to extend the same restriction, or removed (DELETED) if the ``meetings'' feature will no longer be capped by a limit. As with feature management, modifying a rule in this context refers only to structural changes, such as renaming it or adjusting the set of features it affects ---it does not include defining the logic that determines how the rule is evaluated (e.g., setting the actual time limit), which falls under the Evaluation Configuration area.

    \item \emph{Feature dependency management}: In pricing-driven feature toggling, certain features may impose usage-based constraints that depend on whether another feature is enabled in the user’s subscription. To handle this, the system must support feature linking, ensuring that a feature dependent on another is only evaluated if the primary feature is active. For example, in Zoom, ``cloud recording storage'' is available only for PRO and BUSINESS plans. If its storage limit is modeled as a separate toggle, it should not even be evaluated unless ``cloud recording storage'' is enabled for the requesting user. This mechanism not only ensures a clear separation of concerns within the feature toggle structure but also optimizes performance by avoiding unnecessary evaluations.

    \item \emph{Centralized Feature Management}: A foundational characteristic of any feature toggling system is the separation of toggle evaluation logic from the application codebase. This ensures that toggles are not hard-coded within the application logic but instead managed externally, e.g. in a configuration file, database, or dedicated management service (see Fig. \ref{fig:evaluationsInFile}). Centralizing feature management in this way decouples deployment from changes in toggles' evaluation, thereby enabling dynamic updates ---such as pricing ones--- to be performed without requiring code modifications or redeployments.
\end{itemize}
    
    


\subsubsection{Evaluation Configuration}
This area of \modelName focuses on defining the logic that governs feature toggles' activation through rule evaluation expressions. In pricing-driven feature toggling, these evaluations must go beyond basic boolean checks, supporting dynamic expressions that incorporate real-time external data to model pricing constraints and user-specific contexts (see Fig. \ref{fig:evaluationsInFile}).

To meet these requirements, \modelName specifies that evaluation expressions must be both flexible and expressive, allowing for the evaluation of diverse data types (e.g., boolean, numeric, and text) and supporting a set of logical operations. Below, we outline the features required for achieving such expressiveness:



\begin{itemize}[align=left, leftmargin=0pt, labelindent=\parindent,
listparindent=\parindent, labelwidth=0pt, itemindent=!]
    \item \emph{Dynamic evaluation of features}: While separating the evaluation logic of feature toggles from the codebase is a fundamental step, it is insufficient for pricing-driven feature toggling if the configuration only supports static true/false values. Instead, the system must support evaluation expressions that determine feature activation based on conditions evaluated at runtime. 
    
    \item \emph{Boolean values support}: rules' expressions must support boolean variables. For example, a toggle's evaluation controlling whether a Zoom user has access to ``records'' could be configured as: \code{user.subscription.includes("records")}.

    \item \emph{Numeric values support}: the system must manage numeric values within expressions that can be compared using operators such as $>$, $<$, $\geq$, $\leq$, $=$, and $\neq$. For example, Zoom can represent the ``max time per meeting'' restriction with the following expression, which deactivates the ``meetings'' feature once the limit is exceeded:\code{user.currentTime <= user.subscription["maxTimePerMeeting"]}

    \item \emph{Text values support:} evaluation expressions should also support textual comparisons, such as equality ($=$), inequality ($\neq$), regular expression matching, substring checks, or containment operations. For example, Zoom’s ``chat support'' feature could be restricted based on language preferences using an expression like: \code{user.language.includes("EN")}, ensuring that the feature is only available to english-speaking users.

    \item \emph{Dynamic Context-Aware Evaluation}: in pricing-driven feature toggling, evaluation expressions should not only define conditions using static values but also allow for the integration of dynamic, runtime data sourced from external contexts. This enables feature toggles to adapt their behavior based on real-time attributes such as a user’s subscription plan, usage limits, or specific account configurations, rather than relying on pre-defined static conditions. For instance, consider the following hypothetical implementation of the evaluation of the usage limit ``max assistants per meeting'' in Zoom:\\
    \\
      \hspace*{1cm} \code{if (user.plan === "BASIC" || user.plan === "PRO")\{}\\
        \hspace*{2cm} \code{return meeting.assistants <= 100;}\\
      \hspace*{1cm} \code{\}else if (user.plan === "BUSINESS")\{}\\
          \hspace*{2cm} \code{return meeting.assistants <= 300;}\\
      \hspace*{1cm} \code{\}else\{ return false \}}\\

    While functional, this approach scales poorly ---adding an add-on like ``huge meetings'' (which extends the limit to 1000 when contracted) would require further conditional branches, making the logic increasingly complex and harder to maintain. Instead, a dynamic evaluation that directly references external subscription attributes simplifies this logic significantly: \\ 
    \hspace*{0.2cm}\code{return meeting.assistants <= user.subscription["maxAssistants"];
    }\\

    \item \emph{Custom attributes for evaluations}: experience in analyzing feature toggling solutions has showcased that most of them rely on a predefined set of attributes ---such as user ID, IP address, or geographical location--- to configure evaluations \cite{ICSOC2024}. However, in pricing-driven feature toggling, where constraints often stem from highly customizable pricing plans and user-specific conditions, these attributes are insufficient. To address this, feature toggling systems must allow the definition of custom attributes that can be referenced in evaluation expressions. While built-in attributes may still be provided by the system (e.g., default user properties), developers should have the flexibility to define and use additional attributes relevant to their application’s pricing in their evaluations. Moreover, these attributes should support not only primitive data types ---numbers, strings, and booleans--- but also complex structures like objects or arrays, enabling more advanced evaluations.

    \item \emph{Complex logical evaluations}: a pricing-driven feature toggling system must support composite conditions using AND and OR operators to combine multiple evaluation expressions. This allows defining more precise rules, such as requiring a specific plan and add-on to enable a feature. Without this capability, the system would struggle to model complex constraints effectively.
\end{itemize}   

\subsubsection{Feature Evaluation} This area focuses on how feature toggles are queried and utilized at runtime, building on the configuration capabilities established in the previous one. A pricing-driven feature toggling system must provide efficient and flexible mechanisms to evaluate individual or multiple features, handle potential evaluation failures, and ensure consistent output formatting. To accomplish this, \modelName defines the following features for an ideal solution:

\begin{itemize}[align=left, leftmargin=0pt, labelindent=\parindent,
listparindent=\parindent, labelwidth=0pt, itemindent=!]

    \item \emph{Single feature evaluation}: the system must support querying the evaluation of a single feature based on a given user or context parameters. This capability is essential for applications like Zoom, where access to certain server functionalities depends on the user’s subscription. For instance, before granting access to the ``cloud recording'' API, Zoom’s server must verify whether the user performing the request has the plan PRO or BUSINESS in his subscription.

    \item \emph{Multi-feature evaluation}: in many scenarios, retrieving the evaluation results for a set of features at once is more efficient than performing multiple separate queries. This is particularly relevant in the client-side of applications, where a single user operates within a fixed configuration. Instead of evaluating each feature separately, the client can pre-compute and cache the results of all toggles during initialization, reducing redundant evaluations and improving performance. Cached values are updated only when changes in the user’s context require a re-evaluation of a specific subset of features.

    \item \emph{Default values}: to prevent unexpected behavior, the system should allow defining a fallback value when an evaluation fails due to missing data or system errors. This ensures that applications handle failures gracefully instead of displaying inconsistent or unintended behavior.

    \item \emph{Standardized boolean results}: regardless of the complexity of the evaluation conditions or data types involved, the solution should allow forcing the final output of any pricing-driven feature toggle's evaluation to be a boolean value. This guarantees a uniform interpretation of feature availability, simplifying integration across different environments and use cases.

\end{itemize}

\subsubsection{Integration} \modelName defines in this area how a pricing-driven feature toggling system should interact with the applications that integrate it. As any other system, a well-designed integration should abstract the complexity of feature evaluations, making the process seamless for developers. Ideally, from the perspective of the system that uses the feature toggling engine, evaluating a feature should be as simple as calling a function like \textit{isFeatureAvailable(featureId)}, without requiring knowledge of the internal logic governing the toggle. To enable this, \modelName identifies two primary integration approaches: SDK-based integration and API-based integration.

\emph{SDK-based integration} provides applications with direct access to feature evaluations within their execution environment. However, the implementation differs between server-side SDKs and client-side SDKs, as their evaluation strategies must align with their respective operational constraints. Server-side SDKs, which run on backend services processing requests from multiple users, must evaluate each feature dynamically per incoming request. In contrast, client-side SDKs are designed for applications running on user devices (e.g., web or mobile clients), where only a single user’s context is relevant at any given time. As discussed in the previous area, client-side implementations can take advantage of this by pre-evaluating all applicable feature toggles at startup, caching the results, and using them throughout the session ---reducing redundant evaluations and improving performance.

Alternatively, \emph{API-based integration} provides a language-agnostic way to interact with the feature toggling system, enabling applications from different technologies to query feature evaluations remotely. However, this approach must ensure \textit{secure communication and integrity} of both user and pricing information, preventing unauthorized modifications that could lead to illegal feature access.

\subsubsection{Pricing-Driven Automation}. The configuration area is what ultimately distinguishes a feature toggling system as a pricing-driven solution. While the previous areas ensure that feature toggles can be effectively managed and evaluated within a pricing-aware environment, this final area automates their creation and configuration based on the pricing model itself. To achieve this, the system must support a \textit{pricing representation model} that defines the set of features, usage limits, and evaluation rules required for pricing enforcement, such as Pricing2Yaml \cite{CAISEFORUM2024}. By leveraging this model, all relevant feature toggles and their associated conditions can be dynamically generated and managed, eliminating the need for manual setup whenever a pricing update occurs. Nevertheless, to fully enable this approach, the system must provide support for \emph{hot context change management} ---allowing modifications to take effect without requiring application redeployment. This is usually achieved by leveraging custom event definitions.

\section{Validation of Feature Toggling Solutions}
\label{sec:validation}

In this section, we aim to demonstrate the utility of \modelName by applying it to the set of industrial feature toggling tools identified in \cite{ICSOC2024} as having the potential to reach full support for pricing-driven feature toggling, along with the academic solution which indeed achieves it: Pricing4SaaS. Our objective is twofold. First, we aim to verify that our framework produces similar results to those reported in \cite{ICSOC2024}, thereby confirming that \modelName serves as a refined, i.e. more granular, twin of the classification previously proposed. Second, we seek to point out the specific areas where industrial tools fall short in supporting pricing-driven feature toggling, a topic that was not addressed in the earlier study.

To perform the classification, several considerations were taken into account. First, if a solution includes a pricing, we selected the most comprehensive subscription available for that tool. Second, only features accessible through an SDK or an API were considered supported. The core results of our evaluation are summarized in Table \ref{tab:results}, which captures the support provided by each solution across the \modelNameNoSpace's capabilities.

\begin{table}[htp]
    \vspace{-0.2cm}
  \centering
  \begin{adjustbox}{max width=1.0\textwidth}
  \begin{tabular}{|l|c|c|c|c|c||l|c|c|c|c|c|}
    \hline
    \textbf{\modelName Capabilities} & 
      \rotatebox{90}{\href{https://www.getunleash.io/}{\textbf{Unleash}}} & 
      \rotatebox{90}{\href{https://www.devcycle.com}{\textbf{DevCycle}}} & 
      \rotatebox{90}{\href{https://launchdarkly.com}{\textbf{LaunchDarkly}}} & 
      \rotatebox{90}{\href{https://www.togglz.org}{\textbf{Togglz}}} & 
      \rotatebox{90}{\href{https://pricing4saas-docs.vercel.app}{\textbf{Pricing4SaaS}}} &
      \textbf{\modelName Capabilities} &
      \rotatebox{90}{\href{https://www.getunleash.io/}{\textbf{Unleash}}} & 
      \rotatebox{90}{\href{https://www.devcycle.com}{\textbf{DevCycle}}} & 
      \rotatebox{90}{\href{https://launchdarkly.com}{\textbf{LaunchDarkly}}} & 
      \rotatebox{90}{\href{https://www.togglz.org}{\textbf{Togglz}}} & 
      \rotatebox{90}{\href{https://pricing4saas-docs.vercel.app}{\textbf{Pricing4SaaS}}} \\
    \hline
    \multicolumn{6}{|c||}{\textbf{Feature Management}} & \multicolumn{6}{c|}{\textbf{Feature Evaluation}}\\ 
    \hline     
    Feature CREATE  
      & \cmark & \cmark & \cmark & \cmark & \cmark 
      & Single feature evaluation &  \cmark & \cmark & \cmark & \cmark & \cmark \\ 
    \hline
     Feature READ    
      & \cmark & \cmark & \cmark & \cmark & \cmark 
      & Multi-feature evaluation    
       
      & \smark & \smark & \cmark & \xmark & \smark \\
    \hline
     Feature UPDATE  
      & \cmark & \cmark & \cmark & \cmark & \cmark 
      & Default values support & \cmark & \cmark & \cmark & \xmark & \cmark \\
    \hline
     Feature DELETE  
      & \xmark & \cmark & \cmark & \cmark & \cmark 
      & Standardized boolean results & \cmark & \cmark & \cmark & \cmark & \cmark \\
    \hline                 
     Rule CREATE  
      & \cmark & \xmark & \cmark & \cmark & \cmark 
      & \multicolumn{6}{c|}{\textbf{Integration}} \\
    \hline
    Rule READ    
      & \cmark & \xmark & \cmark & \smark & \cmark 
      & Server SDK                  
      & \cmark & \cmark & \cmark & \cmark & \cmark \\
    \hline
    Rule UPDATE  
      & \cmark & \xmark & \cmark & \cmark & \cmark 
      & Client SDK              
      & \cmark & \cmark & \cmark & \xmark & \cmark \\
    \hline
    Rule DELETE  
      & \cmark & \xmark & \cmark & \cmark & \cmark 
      & API-based integration             
      & \cmark & \cmark & \cmark & \xmark & \xmark \\
    \hline
    Feature dependency management 
      & \cmark & \xmark & \cmark & \xmark & \smark 
      & Secure communication        
      & \smark & \cmark & \cmark & \xmark & \cmark \\
    \hline
    Centralized feature management  
      & \cmark & \cmark & \cmark & \cmark & \cmark 
      & \multicolumn{6}{c|}{\textbf{Pricing-Driven Automation}} \\
    \hline
    \multicolumn{6}{|c||}{\textbf{Evaluation Configuration}} 
      & Support of pricing model          
      & \xmark & \xmark & \xmark & \xmark & \cmark \\
    \hline
    Dynamic feature evaluation      
      & \cmark & \cmark & \cmark & \cmark & \cmark
      & Pricing-driven toggle generation      
      & \xmark & \xmark & \xmark & \xmark & \cmark \\
    \hline
    Boolean value support   
      & \xmark & \cmark & \xmark & \cmark & \cmark 
      &  Hot context change management  
      & \cmark & \cmark & \smark & \xmark & \cmark \\ 
    \hline
    Numeric value support   
      & \cmark & \cmark & \cmark & \cmark & \cmark 
      & \multicolumn{6}{c|}{} \\
    \hline
    Text value support      
      & \cmark & \cmark & \cmark & \cmark & \cmark 
      & \multicolumn{6}{c|}{} \\
    \hline
    Context-aware evaluation
      & \cmark & \cmark & \cmark & \cmark & \cmark 
      & \multicolumn{6}{c|}{} \\
    \hline
    Custom attributes for evaluations 
      & \cmark & \cmark & \smark & \cmark & \cmark 
      & \multicolumn{6}{c|}{} \\
    \hline
    Complex logical evaluations  
      & \smark & \smark & \smark & \cmark & \cmark 
      & \multicolumn{6}{c|}{} \\
    \hline
    
  \end{tabular}
  \end{adjustbox}
  \caption{Comparison of feature toggling tools based on \modelName capabilities.}
  \label{tab:results}
  \vspace{-0.7cm}
\end{table}

This evaluation confirms that while existing industrial feature toggling tools are robust in addressing general toggling use cases, they are not inherently equipped to support the variability of pricing-driven environments out of the box. In particular, solutions such as Unleash, DevCycle, LaunchDarkly, and Togglz ---originally designed to facilitate canary releases and A/B testing--- often include functionalities that exceed the minimal requirements for pricing-driven scenarios, yet they fall short in several critical areas. For instance, many of these tools lack comprehensive support for multi-feature evaluation, or restrict logical evaluations to conjunctive (AND) operations. Although disjunctive logic (OR) can be simulated by applying De Morgan’s laws \cite{rosen1999discrete}, this approach complicates configuration, so it doesn't compensate for the absence of native support.

Focusing on each tool, Unleash is one of the most comprehensive solutions; however, it suffers from a significant drawback: features cannot be deleted via API or SDK ---they can only be archived, with full deletion available solely through the UI.  This limitation hinders the implementation of automated, pricing-driven generation of the toggling infrastructure, as features cannot be programmatically removed and re-created. Moreover, Unleash relies on string comparisons for boolean evaluations, which further complicates configuration. Similarly, DevCycle confines rule management to manual operations through its UI, precluding programmatic creation or modification of rules.

LaunchDarkly offers a competitive feature set but presents some limitations. Its boolean evaluations rely on string comparisons (as well as Unleash), and its custom context creation does not allow for defining fixed schemas ---potentially leading to unpredictable behavior in dynamic environments (such as pricing-driven ones). Additionally, its hot context change management is indirect, requiring the development of custom analytics that must be tracked to trigger reloads. Even so, this solution meets all the essential requirements for initiating the implementation of pricing-driven support, as these shortcomings do not significantly hinder its overall viability for the task.

Togglz distinguishes itself with high programmability, permitting the implementation of custom activation strategies, i.e. rules. Yet, this flexibility comes at a cost: evaluation rules are embedded within strategy classes and lack centralized visibility, complicating management. Moreover, since features are defined within an enum, any configuration change necessitates a full application redeployment, making hot context change management impossible.

In contrast, the academic solution, Pricing4SaaS, fully satisfies the criteria outlined by \modelName. It automates the generation of feature toggles from a pricing serialized in Pricing2Yaml (see Section \ref{sec:background:pricing}) and supports hot context change management, because the evaluation engine reads the serialized pricing file on every evaluation ---so any update to the file is automatically reflected without redeployment. Although it currently offers a limited set of SDKs (for \href{https://www.npmjs.com/package/pricing4react}{React}, \href{https://www.npmjs.com/package/pricing4ts}{Node}, and \href{https://central.sonatype.com/artifact/io.github.isa-group/Pricing4Java}{Java}) and does not integrate via API, Pricing4SaaS is purpose-built for pricing-driven feature toggling. Nonetheless, there remains room for improvement, such as enabling bulk evaluation of feature subsets and expanding dependency management between features beyond merely linking usage limits.

In summary, our evaluation validates the results of \cite{ICSOC2024}. Although none of the industrial tools currently supports pricing-driven feature toggling, they achieve the basic level 2 capabilities ---defined by ``dynamic feature evaluation'', ``centralized feature management'', ``context-aware evaluations'', and ``single feature evaluations'' features in \modelName. With the exception of LaunchDarkly ---which already provides all the essential functionalities to extend pricing-driven support--- the remaining tools lack some key features that must be implemented before they can be adapted to pricing-driven environments. Our framework has not only identified these gaps but has also specified the improvements needed to bridge them. Moreover, the classification confirms that Pricing4SaaS is able to manage pricing-driven feature toggling, even though there is room for further enhancement. Hence, these results have confirmed both the utility of \modelName and its ability to serve as a benchmark for future development.




\section{Conclusions and Future Work}
\label{sec:conclusions}

In this paper, we introduced \modelName ---a classification and comparison framework that evaluates the support provided by feature toggling solutions for pricing-driven environments. Its capability-based approach identifies functional gaps during classification, providing a more granular assessment than previous approaches. An initial application to industrial tools confirms prior findings \cite{ICSOC2024}, and reveals LaunchDarkly as the only solution with the essential features to begin developing automatic support for pricing-driven feature toggling. Overall, our results demonstrate that \modelName can be an effective benchmark for developing these systems and highlight the need for targeted improvements in existing industrial tools.

Future work will focus on two key directions. First, we plan to apply \modelName to a broader range of industrial tools and conduct real-world case studies to further leverage its practical applicability. Second, given that Unleash is an open-source solution and ranks as the second most promising tool for pricing-driven feature toggling in our evaluation, we intend to contribute to the project by developing an extension that implements the missing functionalities for pricing-driven automation. This contribution will serve as a stepping stone toward achieving full support for pricing-driven feature toggling in industrial systems.

As a final remark, it is important to recognize that although the concept of pricing traditionally connotes monetary values, the underlying mechanisms of pricing-driven feature toggling extend well beyond cost considerations. The auto-adaptability inherent in these methodologies provides a powerful framework not only for managing dynamic pricings, but also for handling variability in systems that must cater to diverse user segments. For example, the same feature toggling techniques used to adjust feature availability based on subscriptions can be employed to differentiate configurations for various organizational roles or user groups. In such cases, the pricing metaphor is applied without an associated price, leveraging a structured approach to variability and automated adaptation to deliver tailored experiences. This broader applicability underscores the versatility of pricing-driven approaches and highlights their potential to simplify the management of complex, multi-dimensional system configurations.

\section*{Acknowledgments}

This work has been partially supported by grants 
PID2021-126227NB-C21, and 
PID2021-126227NB-C22      
funded by MCIN/AEI/10.13039/501100011033/FEDER and European Union ``ERDF a way of making Europe'';
and %
TED2021-131023B-C21 and 
TED2021-131023B-C22     
funded by MCIN/AEI/10.13039/501100011033 and European Union ``NextGenerationEU''/PRTR.

\bibliographystyle{splncs04}
\bibliography{references}

\end{document}